# Vulnerability Prediction Based on Weighted Software Network for Secure Software Building

Shengjun Wei*
School of Computer Science & Tchnology
Beijing Institute of Technology
Beijing, China
sjwei@bit.edu.cn

Hao Zhong
School of Computer Science & Tchnology
Beijing Institute of Technology
Beijing, China
zhonghao@bit.edu.cn

Chun Shan
School of Computer Science & Tchnology
Beijing Institute of Technology
Beijing, China
sherryshan@bit.edu.cn

Lin Ye
School of Computer Science and Technology
Harbin Institute of Technology
Harbin, China
hityelin@hit.edu.cn

Xiaojiang Du
Department of Computer and Information Sciences
Temple University
Philadelphia, PA, USA
dxj@ieee.org

Mohsen Guizani
Department of Electrical and Computer Engineering
University of Idaho
Idaho, USA
mguizani@ieee.org

*Abstract*—To build a secure communications software, Vulnerability Prediction Models (VPMs) are used to predict vulnerable software modules in the software system before software security testing. At present many software security metrics have been proposed to design a VPM. In this paper, we predict vulnerable classes in a software system by establishing the system's weighted software network. The metrics are obtained from the nodes' attributes in the weighted software network. We design and implement a crawler tool to collect all public security vulnerabilities in Mozilla Firefox. Based on these data, the prediction model is trained and tested. The results show that the VPM based on weighted software network has a good performance in accuracy, precision, and recall. Compared to other studies, it shows that the performance of prediction has been improved greatly in Pr and Re.

*Keywords—Secure Communication Software, Vulnerability Prediction, Software Metrics, Weighted Software Network*

## I. INTRODUCTION

Secure software is a core requirement of many modern information systems [1]. However, the security problems of these systems are more and more prominent. From wireless sensor network widely used in military field previously [2,3,4] to modern medical equipment [5], there are many reports about successfully exploiting vulnerabilities to break these systems. During software products developing, due to the time and budget constraints, the developers can not find out all the vulnerabilities. If a developer can be aware of the vulnerability-prone modules in these software systems, and prioritize the limited test resources on these modules, more vulnerabilities would be found. In paper [6], vulnerability prediction models (VPMs) are firstly proposed to meet this requirement. The VPMs are established by using machine learning algorithms, the output of VPM is vulnerability-proneness of a software module [7-10].

Compared with the other vulnerability analysis techniques (e.g., [11,12]), VPM is used as a guidance before security testing. At present, a lot of VPMs have been proposed [6-10, 11-19]. Software metrics are utilized as features in these VPMs, and the training data are obtained from public vulnerabilities database. The ways of VPM research refer to software defect prediction models (DPMs). Some studies [1,20] believe that high complexity, high coupling, and low cohesion can lead to the difficulty of software understanding, developing, testing and maintaining and may introduce vulnerabilities into software systems. The complexity describes the complex degree in a module, the coupling describes the relevancy between two modules and the cohesion describes the relevancy between two members in a module. So the dependencies among modules in a software system, as well as dependencies among members in these modules, can be used as features to predict the vulnerability-proneness of a module [9,10].

In a software network [21], a node is a code unit, it could be a class, a method, a package, a component，a binary, or even a subsystem, and an edge is the dependency between two code units. So the dependencies among software modules are represented graphically by the software network. Paper [22] use network analysis to predict defects based on software networks. Paper [9] use nodes and edge attributes in dependency graphs as features to predict vulnerable software components. The software networks in [22] and [9] are Unweighted network, but the edges of a software networks should be weighted because the degree of dependence between two nodes is not just 0 or 1, the degree should be between [0,1]. Paper [23] demonstrates how to establish a weighted software network and analyze the relationship between the attributes of networks and the defects. It was discovered that the attributes of networks are significantly associated with defects. In our study, we establish a weighted software network at class granularity, and utilize the attributes of the weighted network to predict the vulnerable classes.

The rest of this paper is organized as follows: Section 2 explain the reason of using a weighted software network to build a VPM, while Section 3 provides related work. Section 4 introduces the weighted software network and metrics. Section 5 describes the case study design and evaluation criteria. Section 6 provides the results and discusses our findings and limitations. Finally Section 6 summarizes our study.

## II. MOTIVATION

The research of software network discovered that the defect-proneness or vulnerability-proneness of a node in the software network is associated with the attributions of the network, and is summed up as follows:

1) Paper [22] builds a software network targeting Windows Server 2003, in which the node is binary, and they discovered two phenomena: one is that central binaries tend to be defect-prone, and the other is that the larger a clique, the more defect-prone are its binaries. In the first circumstance, a typical network motif is a star pattern, the center of the star tend to be defect-prone. In the second circumstance, larger clique tend to be defect-prone. A clique is a set of binaries for which between every pair of binaries a dependency exists and a clique is maximal if no other binary can be added without losing the clique property.

2) Paper [9] builds a software network, in which a node is a file, and they conducted experiments on JSE, they find that the vulnerability-proneness of a node is associated with the node's in-degree and out-degree, and then they build a VPM to predict vulnerable files.

3) Paper [23] builds a software network, in which a node is a class, and the network is weighted and directed. They study the relationships between the intensity of a node and the node's ability of defects propagation and aggregation. They found and validated that the larger the intensity of a node, the stronger the ability of defects propagation and aggregation. Furthermore, high-intensive nodes include most of the pre-testing defects.

Based on the conclusions that the vulnerability-proneness of a software unit is associated with the attributions of its software network, we attempt to adopt a weighted software network to predict the vulnerable class. In this paper, we illustrate that in comparison to the traditional unweighted and undirected software network, the weighted and directed software network is more suitable to depict the dependencies among software units. Then we build a weighted software network at class level, and extract several typical attributions of network to establish a VPM.

## III. RELATED WORK

The software metrics which includes code complexity and developing process metrics which includes code churn, developer's experience and the development team's organizational structure, etc. are used to build Defect Prediction Models (DPMs) to predict potential defect-proneness of program modules [24-26], and then by reference, these metrics are used to build VPMs because of regarding software security vulnerabilities as special defects.

Y. Shin et al.[13,14] analyzed the relationship between nine traditional complexity metrics and security vulnerabilities, and then established the prediction model. They conducted an experiment targeting JSE which showed that the models have a high false negative rate. They then included complexity metrics of software design and program execution in their model, and results show that the false negative rate is reduced [7]. In paper [8], Y. Shin et al. built prediction model with complexity, code churn and fault history metrics, experiments targeting Mozilla Firefox show that the model has a recall rate higher than 80%, but the false positive rate is higher than 20%. Then, in paper [15], they used complexity, code churn, and developer activity metrics to predict vulnerable files, experiments targeting Mozilla Firefox and Red Hat Linux kernel show that the model can predict over 80 percent of the known vulnerable files with less than 25 percent false positives for both projects. T. Zimmermann et al [16] used code churn, code dependencies, and organizational metrics to build VPMs. They conducted experiments on binary files in windows vista that resulted models having high precision but low recall.

VH Nguyen et al [9] used complexity and dependency metrics, which are obtained based on a component dependency graph defined by themselves, to establish a VPM. This dependency graph is a kind of a software network graph. They conducted experiments on JSE which resulted in the model having good accuracy and false positive rate but lower recall rate. I. Chowdhury et al [10] used traditional complexity, coupling, and cohesion metrics together to establish a VPM. They conducted experiments on Mozilla Firefox, and collect all the public vulnerabilities of 52 releases up to the experiment date. They concluded that the complexity, cohesion, and coupling metrics are efficient for vulnerability prediction, and that these three types of metrics collectively affect vulnerability-proneness of a software unit rather than one single category of metrics. S. Neuhaus et al [6] discovered that the vulnerability-proneness of a file is connected with the import/function calls in the file through an association rule mining algorithm. Then they used import/function calls to establish a classifier, with Firefox as experiment object, the model has a recall rate of 45% and an accuracy of 70%.

By considering a source file as a text, R. Scandariato et al used a text mining technique to predict vulnerable file [17] by taking a word in the text as a feature to establish a classifier. In their following study, they compared this text mining method with software metrics method based on the same vulnerability database, and discover that text mining has a higher recall rate [18]. M. Jimenez et al [19] also compared the performance among three methods: text mining, software metrics and import/function calls, and they discovered that the software metrics method is the worst.

At present, the researches of VPM includes three aspects: the more efficient metrics, the more powerful machine learning algorithm, and the higher-quality vulnerability database. According to the public materials about this area, the performance of VPM should be improved substantially. In this paper, we propose some new metrics to establish the VPM.

## IV. SOFTWARE NETWORK AND THE METRICS

A software network presents the architecture of a software system graphically. At different granularities, the nodes of a software network could be component, package, subsystem, class, method, and data etc., and the edges of which are the dependencies among the nodes. In our study, we establish a weighted software network, in which a node is a class.

### A. Weighted Software Network

Definition 1. The weighted software network of a software system is a weighted direct graph $WSN=(V,E,W)$, where: $V$: is a set of nodes. A node represents a class; $E$: is a set of directed edges. When there exits dependencies between two classes, there is a edge between the two corresponding nodes. The direction of a edge is point to the dependent object. The dependencies include inheritance and association relations, etc; $W$: is a degree of dependency for an edge, which belong to [0,1]. 0 is no dependency, and 1 is a complete dependency. The calculation formula is depicted in Definition 2.

Definition 2. Supposing $Vf_i$ is the sum of functions in a node $i$, $Vf_{ij}$ is the sum of dependent functions from node $i$ to node $j$ in the node $i$, $Vf_j$ is the sum of functions in a node $j$, and $Vf_{ji}$ is the sum of functions in the node j being depended on by node i. The value of $w_{ij}$ is calculated as:

$$w_{ij}=(Vf_{ij}/Vf_i)\times(Vf_{ji}/Vf_j) \qquad (1)$$

Obviously, The degree of dependency between two classes is associated with the numbers of dependent functions in each classes. For example, as shown in Fig. 1(a), class A has three functions, which are a(), x(), b(). The three functions are all dependent on one function c() in class B. In Fig. 1(b), class A has just one function a() which is dependent on one function c() in class B. In this figure, it is clear that the degree of dependency between class A and B in Fig. 1(a) and Fig. 1(b) is not equal, Fig. 1(a) is greater than Fig. 1(b). However, the traditional software network of Fig. 1(a) and Fig. 1(b) can not depict this situation, as shown in Fig. 1(c), they are the same. If we establish the weighted network by calculating the weights with the definition 2, as shown in Fig. 1(d) (which is corresponding to Fig.1(a)) and Fig. 1(e) (which is corresponding to Fig.1(b)), the degree of dependency in each situation is not equal, and it is more suitable for depict the dependency in the situation.

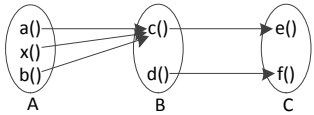

(a). The dependency between class A and B: three functions in class A are dependent on one function in class B

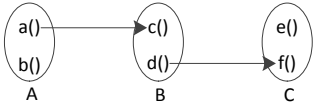

(b). The dependency between class A and B: One function in class A is dependent on one function in class B

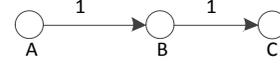

(c). The traditional software networks for situation (a) and (b): they are the same

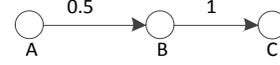

(d). The weighted software network for situation (a)

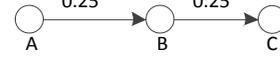

(e). The weighted software network for situation (b)

Fig. 1. Weighted Software Network

### B. Metrics

Paper [23] discovers that the number of defects and the defects propagation ability of a node are associated with the attributes of the software network significantly. If the weights of input or output are larger, the defect-proneness is larger. Similarly, paper [9] discovers that a component's vulnerability-proneness is associated with the component node in the component dependency graph, but the graph in [9] is not weighted and directed. Our metrics are defined based on a weighted software network at class level to predict the vulnerability-proneness of a class. The metrics are shown in table 1.

TABLE I. METRICS BASED ON WEIGHTED SOFTWARE NETWORK

| Metrics | Description and rationale |
|---|---|
| NumofLn | Total number of lines of code in a node. The larger the value, the higher the complexity of the class. |
| NumofFn | Number of functions in a node. The larger the value, the higher the complexity of the class. |
| AveCCofFn | Average Cyclomatic Complexity of all the functions in a node. The larger the value, the higher the complexity of the class. |
| IntofIn | We define the IntofIn as the intensity of input of a node, it is corresponding to the degree of input in a unweighted network. The IntofIn of node $i$ is sum of weights of all the input edges of node $i$. The IntofIn reflects the dependency from the other node to node $i$, the larger the value, the more important the node $i$, and the higher the coupling between node $i$ and the other nodes. |
| IntofOut | We define the IntofOut as the intensity of output of a node, it is corresponding to the degree of output in a unweighted network. The Intofout of node $i$ is sum of weights of all the output edges of node $i$. The IntofOut reflects the dependency from node $i$ to the other node, the larger the value, the more important the node $i$, and the higher the coupling between node i and the other nodes. |
| ClusCoeofNode | We define ClusCoeofNode as clustering coefficient of a node $i$: supposing node $i$ has $k_i$ adjacent nodes and there have $M_i$ edges among these $k_i$ adjacent nodes, the ClusCoeofNode is equal to the sum of all $M_i$ edges divided by $k_i(k_i-1)$. ClusCoeofNode reflects the cluster degree of a class in a software system. The larger the value, the higher the cluster degree, and the more important the node. |
| BetwofNod | BetwofNod is the betweenness of a class node. There has at least one path between any two reachable nodes in the |

| Metrics | Description and rationale |
|---|---|
| | network. The shortest path is the path with the smallest sum of weights of all the edges in the path. The BetwofNod of node i is the number of shortest paths across the node *i*. The larger the value, the more important the node. |

## V. CASE STUDIES

We conducted experiments on Mozilla Firefox, an open source software system. The vulnerabilities of Firefox are published by Mozilla Foundation Security Advisories (MFSAs) on a public website [27]. For the releases before FireFox 43, the MFSAs can be accessed freely, but need authorization for the releases after FireFox 43. We collected all the vulnerabilities of the releases from the first to FireFox 43. For Mozilla Firefox, all the commits for bug fix are recorded in Bugzilla, so the patches for mitigating the vulnerabilities in a file can be found in Bugzilla, and the corresponding vulnerable files can also be retrieved. Therefore, the historical vulnerable classes of Firefox can be collected.

### A. Vulnerability Data Collection

There are three steps for collecting vulnerabilities data: 1) Searching vulnerabilities from MFSA, 2) For every vulnerability obtained in step 1, found out the corresponding bug links in Bugzilla, 3) For every bug obtained in step 2, found out the patched files for fixing the bug, and locate the vulnerable classes, then plus 1 to the vulnerability number of the classes. In order to collect these vulnerabilities data automatically, we designed and implemented a crawler tool. The flow chart of the crawler is shown in Figure 2. There are 4 loops.

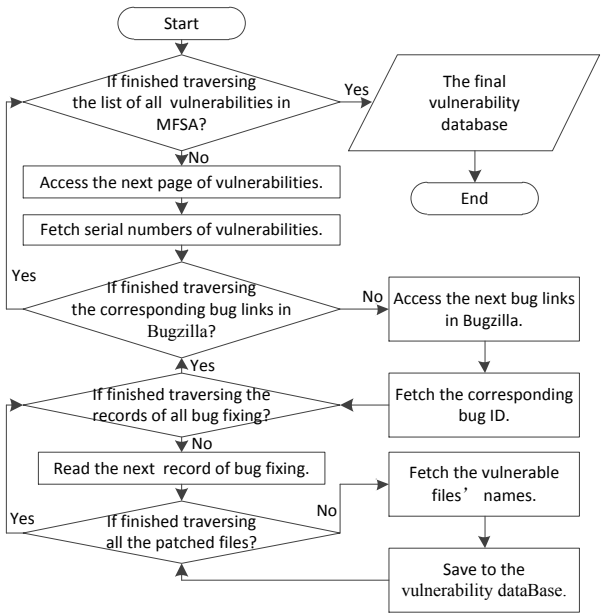

Fig. 2. Flowchart of the crawler tool

Our crawler shows Firefox 43 contains a total of 6,835 C++ classes, and among them, 616 have vulnerabilities. Among the vulnerable classes, there are 378 classes with 1 vulnerability, 98 classes with 2 vulnerabilities, 59 classes with 3 vulnerabilities, 45 classes with 4 vulnerabilities, 22 classes with 5 vulnerabilities, and 14 classes with more than 5 vulnerabilities. During the collection of the vulnerability data, it is assumed that the same vulnerability exists in all previous versions.

### B. Data Collection

We use Understand c++ [28], a commercial tool, to analysis the source code. We develop a software network generation tool based on the Understand c++. We design the required functions. We analyze the code structure, extract the private and static members of a file, fetch dependency information for these members, count and analyze the functions in classes, generate the weighted software network at class level, and then compute the metric values associated with the nodes and edges based on the software network graph. Table 2 is the statistical data of the nodes and edges in the weighted software network.

TABLE II.  STATISTICAL DATA OF FIREFOX 43

| Units | Classes | Functions |
|---|---|---|
| Amount of units (nodes) | 6835 | 9374 |
| Amount of dependences (edges) | 7297 | 14728 |

### C. Prediction Performance Measures

In our experiments, to compare with the result in [9], we select five same machine learning techniques used in [9], they are Bayesian Network(BN), Naïve Bayes (NB), Neural Network(NN), Random Forest(RF), and Support Vector Machine (SVM).

In our prediction model, we classify a class as vulnerable if the output of the model is more than 0.5, otherwise it is not vulnerable. We use four metrics to evaluate the performance of the model. These four metrics are Acc (Accuracy), Re(Recall), FP(False Positive rate), and Pr(Precision). We use a confusion matrix (as shown in table 3) for the binary classification to define each performance measures.

TABLE III.  CONFUSION MATRIX

| Actual | Predicted as | |
|---|---|---|
| | *vulnerable* | *not vulnerable* |
| not vulnerable | TN=True Negative | FP=False Positives |
| vulnerable | FN=False Negatives | TP=True Positives |

The calculation formulas are as follows: Acc=(TP+TN)/(TP+FP+TN+FN), Pr=TP/(TP+FP), FP=FP/(FP+TN), Re=TP/(TP+FN).

## VI. RESULTS AND DISCUSSIONS

### A. Predictive Power

First, We use the Wilcoxon rank sum test to evaluate the discriminative power of the metrics in table 1. The null hypothesis is that there is no statistically significant difference between the metric values of vulnerable classes and non-vulnerable classes. The results showed that the measures of all seven metrics for the vulnerable classes and the non-vulnerable classes in Firefox are significantly different at the 0.05 significance level.

We use a free tool Weka [29] to implement the five machine learning techniques with the parameters initialized with the default settings. The amounts of vulnerable classes in Firefox 43 is less than 10% of the total files. Such a dataset is heavily unbalanced. In this paper, we use under-sampling to balance the dataset. With under-sampling, all the vulnerable classes in the dataset are retained, while only a subset of the non-vulnerable classes are selected. The sample of non-vulnerable classes is randomly chosen that the number of vulnerable classes matches the number of non-vulnerable ones. The final dataset has 1232 items, and the ratio of positive and negative sample is one to one. On the balanced dataset, we use 10×10 cross-validation to train and test the model. By using stratified sampling, the vulnerability dataset is split into 10 folds of equal size, each has 220. The number of positive sample and negative sample are 110 respectively. We use one fold for testing and the other 9 folds for training, rotating each fold as the test fold. The entire process is then repeated 10 times and the results are averaged, as shown in table 4.

TABLE IV. RESULTS IN EXPERIMENT

| Technique | Acc(%) | Pr(%) | FP(%) | Re(%) |
|---|---|---|---|---|
| BN | 84.45 | 84.98 | 14.82 | 83.73 |
| NB | 85.59 | 86.40 | 13.45 | 84.64 |
| NN | 90.50 | 90.83 | 9.18 | 90.18 |
| RF | 84.91 | 85.99 | 13.64 | 83.45 |
| SVM | 84.64 | 86.05 | 13.55 | 82.82 |
| Average | 86.02 | 86.85 | 12.93 | 84.96 |

B. *Discussion*

- From the experiment results, the performance of five techniques are similar, but NN is the best with the highest value of all the four performance measures. Acc, Pr, and Re are all over 90%, and the FP is less than 10%.

- From the average value of five results, Acc, Pr, and Re are all around 85%, and FN is 12.93%. That is, among all the 1232 sample classes, 86.02% are classified correctly, and among the classes classified as vulnerable, 86.85% are actual vulnerable ones. Among all the 616 vulnerable classes, 84.96% can be classified correctly. Among all the non-vulnerable sample classes, 12.93% are classified as vulnerable falsely. These results indicate that the proposed metrics are effective for vulnerable classes prediction.

- To compare with the results of [9], we display side-by-side results adopted from work of [9] and our results in table 5. Table 5 list four kinds of performance values of five machine learning techniques, and Fig. 3 is the histograms, in which, a), b), c), d), e), f) are corresponding to the machine learning techniques BN, NB, NN, RF, SVM, respectively. and g) is corresponding to the average of five. By analyzing the individual or the average results, there is a little increase in the FP, but the Pr and Re have improved a lot, as shown in Fig. 3.g), Pr is increased by 27.7%, Re is increased by 42.7%, and Acc has also increased simultaneously. However, there may be some reasons that have affected this comparison: 1) our weighted software network is at a class level, but the CDG in [9] is at a component level, and 2) we conduct experiments targeting on the Firefox, but [9] just aiming at JSE in Firefox, and our sample database includes more vulnerability data than [9].

TABLE V. RESULTS COMPARISON WITH [9]

| | ACC(%) | | Pr(%) | | FP(%) | | Re(%) | |
|---|---|---|---|---|---|---|---|---|
| | *[9]* | *ours* | *[9]* | *ours* | *[9]* | *ours* | *[9]* | *ours* |
| BN | 81.86 | 84.45 | 55.17 | 84.98 | 16.15 | 14.82 | 74.42 | 83.73 |
| NB | 85.78 | 85.59 | 65.22 | 86.40 | 9.94 | 13.45 | 69.77 | 84.64 |
| NN | 85.29 | 90.50 | 72.41 | 90.83 | 4.97 | 9.18 | 48.84 | 90.18 |
| RF | 84.31 | 84.91 | 62.22 | 85.99 | 10.56 | 13.64 | 65.12 | 83.45 |
| SVM | 85.78 | 84.64 | 85.00 | 86.05 | 1.86 | 13.55 | 39.53 | 82.82 |
| Average | 84.61 | 86.02 | 68.01 | 86.85 | 8.70 | 12.93 | 59.53 | 84.96 |

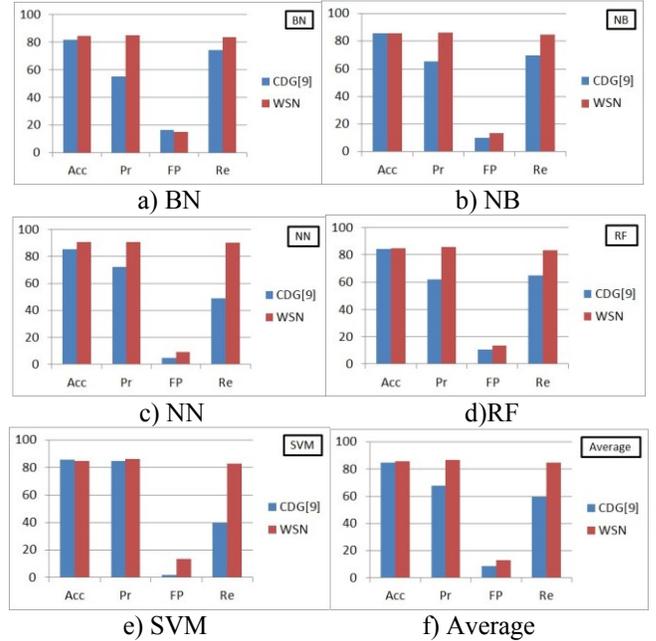

a) BN    b) NB
c) NN    d) RF
e) SVM   f) Average

Fig. 3. Comparison of our results with results in [9]

- A security vulnerability can cause severe damage to an organization, we think we should pay more attention on the Recall rate, but relax FP rate within a rational scope. We believe that it is more important to identify the vulnerable classes, even at the expense of incorrectly predicting some non-vulnerable classes as vulnerability-prone. On the other hand, our average FP rate is 12.93%, still below 15%.

## VII. SUMMARY AND FUTURE WORK

To build a secure software, VPMs are used to predict vulnerable modules in these software systems before security testing. Public researches show that the software network graph can use to establish a VPM. The software networks adopted in these proposed VPMs are unweighted and undirected. In the actual situation of software network application, it should be considered that the weights of the edges would be different. In this paper, we establish a weighted software network at class level and utilize the attributes of the network graph to predict vulnerable classes. We conducted experiments to validate the efficiency of the model and we discovered that the proposed metrics are effective. In our future work, we will conduct more experiments targeting more communication software projects and make comparison with other different works.

ACKNOWLEDGMENT

This work was supported by the National Key R&D Program of China (Grant no. 2016YFB0800700) and the National Natural Science Foundation of China (Grant no. U1636115).